\documentclass[preprint,
superscriptaddress,
nofootinbib,
 amsmath,amssymb,
 aps,prd
]{revtex4-1}

\usepackage{graphicx}
\usepackage{dcolumn}
\usepackage{bm}

\begin{document}

\preprint{RESCEU-19/18}

\title{Gravitational particle creation for dark matter and reheating}

\author{Soichiro Hashiba}
	\email{sou16.hashiba@resceu.s.u-tokyo.ac.jp}
	\affiliation{Department of Physics, Graduate School of Science, The University of Tokyo, Tokyo 113-0033, Japan}
	\affiliation{Research Center for the Early Universe (RESCEU), Graduate School of Science, The University of Tokyo, Tokyo 113-0033, Japan}
\author{Jun'ichi Yokoyama}
	\email{yokoyama@resceu.s.u-tokyo.ac.jp}
	\affiliation{Department of Physics, Graduate School of Science, The University of Tokyo, Tokyo 113-0033, Japan}
	\affiliation{Research Center for the Early Universe (RESCEU), Graduate School of Science, The University of Tokyo, Tokyo 113-0033, Japan}
	\affiliation{Kavli Institute for the Physics and Mathematics of the Universe (Kavli IPMU), WPI, UTIAS, The University of Tokyo, 5-1-5 Kashiwanoha, Kashiwa 277-8583, Japan}

\date{\today}

\begin{abstract}
The purely gravitational dark matter (PGDM) which interacts with the standard model particles only by gravitational interaction has recently been discussed. Due to its feeble interaction, PGDM may be produced mainly by the gravitational particle creation, which plays an important role in the reheating after kinetically driven inflation and some potential-driven inflation without subsequent inflaton oscillating phase. Therefore, we consider the possibility of the gravitational reheating model which can also explain the present PGDM density at the same time. We consider a model where two massive scalar fields are incorporated into the standard model besides the inflation sector. We show that the gravitational particle creation prevails over the thermal creation --- the freeze-in process --- and it can actually explain the reheating and the present abundance of dark matter if one of the scalar particles is as heavy as the Hubble parameter during inflation $\sim10^{13}$ GeV and finally decays into radiation via sufficiently weak coupling, and the other is a stable PGDM with its mass of the order of $10^3$ GeV.
\end{abstract}

\pacs{Valid PACS appear here}
\maketitle


\section{\label{intro}Introduction}
Although the $\Lambda$CDM model explains the evolution of the universe quite well, the identity of cold dark matter (CDM) and its production mechanism are still missing pieces in modern cosmology. According to the results of the cosmic microwave background (CMB) observation \cite{Aghanim:2018eyx}, the dark matter occupies about $27\%$ of the total energy density of the universe. It must not interact with the standard model (SM) particles through electromagnetic interactions since the dark matter is ``invisible". Therefore, it is natural to consider the weakest interacting dark matter which interacts with the SM particles only gravitationally \cite{Chung:1998zb,Chung:1998ua,Kuzmin:1998kk,Chung:2001cb}. Such kind of dark matter is called the Planckian interacting dark matter (PIDM) \cite{Garny2016} and also as the pure(ly) gravitational dark matter (PGDM) \cite{Tang:2016vch,Ema2018}. We use the latter terminology hereafter. PGDM is difficult to create due to its extremely weak interaction.

The reheating process after inflation is also a remaining problem in cosmology. The inflationary universe model (See {\it e.g.}\ Ref.\ \cite{Sato:2015dga} for a review of cosmic inflation.) has amazingly succeeded in explaining the very early universe and the later structure formation by providing an appropriate initial condition of primodial fluctuations, however, how to realize the reheating process --- the particle creation after inflation which turns the inflaton dominated universe into the radiation dominated --- has not been completely understood yet. Inflationary models are roughly classified into two types: those driven by a potential energy of the inflaton and those driven by its kinetic energy. In most potential-driven models \cite{Sato1981,Guth1981,Linde1982,Linde1983}, the reheating can realize by the inflaton field oscillation \cite{Abbott1982,Dolgov1982}. In kinetically driven models \cite{Picon1999,Kobayashi2010} and some potential-driven models \cite{Peebles1999}, the inflaton field does not oscillate after inflation but kination --- the epoch when the kinetic energy of a free scalar field dominates the cosmological energy density --- follows inflation and thus the reheating model mentioned above is not valid. Therefore, it has been claimed that the reheating process is achieved by the gravitational particle creation in these inflationary models \cite{Picon1999,Kobayashi2010,Peebles1999,Kunimitsu2012}. \footnote{The reheating through direct interactions between the inflaton and a matter field preserving shift symmetry has been discussed in Ref.\ \cite{BazrafshanMoghaddam:2016tdk}.}

It is known that any kind of particles which are not conformally invariant is created when the time dependence of the cosmic expansion changes \cite{Parker1969,Zeldovich1971}. This process is called the gravitational particle creation. Although it has usually been analyzed perturbatively for massive scalar particles \cite{Ema2018,Birrell1980,Turner1988,Chung2001,Chung:2018ayg}, in Ref.\ \cite{Hashiba2018} we have calculated the gravitationally produced energy density non-perturbatively. The gravitational particle creation is suitable for the production of PGDM and also for the reheating after inflation if the inflation field oscillation does not follow, (while it is negligible in usual potential-driven inflationary models because it is much less efficient than the decay of coherent inflaton oscillations). Therefore, we examine whether or not the gravitational particle creation can explain the reheating and the production of PGDM at the same time in inflationary models which kination follows inflation such as Refs.\ \cite{Picon1999,Kobayashi2010,Peebles1999}.

This paper is organized as follows. Our previous calculation of the gravitational particle creation \cite{Hashiba2018} is briefly reviewed in Sec. \ref{gpc}. The present abundance of gravitationally produced PGDM is calculated in Sec. \ref{rh}. PGDM should be produced sufficiently in order to prevent gravitons from disturbing CMB. This issue is discussed in Sec. \ref{cc}. Since the freeze-in process also produces PGDM, the gravitational particle creation is compared with it in Sec. \ref{fi}. Our results are summarized in Sec. \ref{concl}. We use the natural units $c = \hbar = 1$ and $M_{\rm Pl} = 1.2\times10^{19}\,$GeV throughout the paper.

\section{\label{gpc}Gravitational particle creation after inflation}
We consider inflationary models in which the kination stage follows inflation such as $k$-inflation \cite{Picon1999} and kinetically driven G-inflation \cite{Kobayashi2010}, and adopt the adiabatic vacuum as the basis of counting the number of particles. The concept of the adiabatic vacuum is that the mode function of the vacuum should approach the positive frequency mode in the Minkowski spacetime since for high-momentum particles the universe looks as if it is almost flat and static for the relevant short time and length scales. The equation of motion of a mode function for a scalar field conformally coupled to gravity in a spatially flat Friedmann-Lema\^{i}tre-Robertson-Walker (FLRW) metric is
\begin{equation}
	\frac{d^2 \chi_k(\eta)}{d\eta^2} + \left[ k^2 + a^2(\eta)m^2 \right] \chi_k(\eta) = 0, \label{eom}
\end{equation}
where $\eta$ and $a(\eta)$ denote the conformal time and the scale factor, respectively. During de Sitter inflation and kination, the scale factor is asymptotically proportional to $(-\eta)^{-1}$ with $\eta<0$ and $\eta^{1/2}$ with $\eta>0$, respectively. Equation (\ref{eom}) can be analytically solved with this scale factor, and the adiabatic vacuum is obtained by imposing the condition
\begin{equation}
	\lim_{k \to \infty} \chi_k(\eta) = \frac{1}{\sqrt{2k}}e^{-ik\eta}
\end{equation}
on the analytic solutions of Eq.\ (\ref{eom}). The adiabatic vacuum can be written in terms of Hankel functions. We have given an exact form in Ref.\ \cite{Hashiba2018}. The produced number density is obtained from the Bogoliubov coefficient of the transformation between this adiabatic vacuum and the mode function of the universe.

Assuming a smooth transition from inflation to kination with time scale $\Delta t$, the energy density of produced particles is given by \cite{Hashiba2018}
\begin{equation}
	\rho = \mathcal{C} e^{-4m\Delta t}m^2 H_{\rm inf}^2, \label{rho}
\end{equation}
where $\mathcal{C} \simeq 2\times10^{-4}$, $m$ and $H_{\rm inf}$ denote the mass of the created scalar particle and the Hubble parameter during inflation, respectively. As seen in Eq. (\ref{rho}), the mass threshold above which the gravitational particle creation is exponentially suppressed is given by the inverse transition time scale $(\Delta t)^{-1}$ rather than the Hubble parameter during inflation $H_{\rm inf}$. Since its power spectrum has a peak around $k \sim m$, the heaviest produced particle soon becomes non-relativistic by the subsequent expansion of the universe.

\section{\label{rh}Reheating and gravitational production of dark matter}
We incorporate two massive scalar particles, both of which are conformally coupled to gravity. Only their mass terms violate the conformal symmetry and serve as a source of gravitational particle production. One of them decays into radiation and realizes reheating, and the other is a stable PGDM. We call them $A$ and $X$, respectively. Several papers consider an ${\rm U}(1)$ interaction between PGDMs \cite{Dubrovich:2003jg,Garny:2018grs}, but here we do not assume any interaction between $X$'s other than gravitational interaction. We assume the decay of $A$ and its decay width given as
\begin{equation}
	\Gamma = \alpha m_A,
\end{equation}
where $\alpha$ is a dimensionless constant. For example, $\alpha$ takes $\lambda^2/32\pi^2$ when $A$ decays into a Fermion pair via a Yukawa-interaction term $\lambda A \bar{\Psi}\Psi$. Since $H \propto a^{-3}$ during kination, the scale factor at $t = t_d$ when $H = \Gamma$ is
\begin{equation}
	a_d = \alpha^{-1/3} \left( \frac{m_A}{H_{\rm inf}} \right)^{-1/3}. \label{ad}
\end{equation}
Here and hereafter we put the scale factor at the end of inflation to unity. Although $A$ constantly decays into radiation until $t_d$, the energy density of $A$ is diluted as $a^{-3}$ more slowly than that of radiation as $a^{-4}$ and then it is enough to consider decay of $A$ around $t = t_d$ where resultant radiation is least diluted. Here, we assume that $A$ decays during kination. If $A$ does not decay until kination ends, the scale factor at the end of kination is
\begin{equation}
	a_{\rm MD} = 1.0 \times 10^5 e^{\frac{4}{3}m_A \Delta t} \left( \frac{m_A}{10^{13} {\rm GeV}} \right)^{-2/3}. \label{aMD}
\end{equation}
If $a_d < a_{\rm MD}$, then $A$ actually decays during kination. According to Eqs.\ (\ref{ad}) and (\ref{aMD}), this condition is satisfied if $\alpha > \mathcal{O}(10^{-17})$ when $m_A \simeq \Delta t^{-1} \simeq H_{\rm inf} \simeq 10^{13}\, {\rm GeV}$. Since almost all of the cosmic entropy is generated by the decay of $A$, the ratio of the energy density of $X$ to the entropy density is conserved after $A$ decay. According to Eqs.\ (\ref{rho}) and (\ref{ad}), the energy density of $X$ at $t = t_d$ is
\begin{equation}
	\rho_X|_d = \mathcal{C}\alpha e^{-4m_X \Delta t} m_A m_X^2 H_{\rm inf}. \label{rhodecay}
\end{equation}
and the entropy density is
\begin{equation}
	s|_d = \frac{2\pi^2}{45} g_{\ast d} T_d^3, \label{sd}
\end{equation}
where $g_{\ast d}$ and $T_d$ denote the effective degrees of freedom, which we take the standard value $106.75$, and the temperature at the time $A$ decays, respectively. According to Eqs.\ (\ref{rho}) and (\ref{ad}), the latter is given by
\begin{equation}
	T_d = 5 \times 10^{-2} \alpha^{1/4}e^{-m_A \Delta t} m_A^{3/4}H_{\rm inf}^{1/4}, \label{Td}
\end{equation}
and hence, the ratio of $\rho_X$ to $s$ is given by
\begin{equation}
	\frac{\rho_X}{s} = 4 \times 10^{-2} \alpha^{1/4} e^{(3m_A-4m_X) \Delta t} \frac{m_X^2 H_{\rm inf}^{1/4}}{m_A^{5/4}}. \label{rhos}
\end{equation}
This should be equal to $\approx 4 \times 10^{-10}\,$GeV in order to explain the present dark matter density \cite{Aghanim:2018eyx}.

\section{\label{cc}Concealing graviton}
The gravitational particle creation mechanism also produces the graviton, whose abundance is twice as much as that of a massless minimally coupled scalar particle. Since the graviton is also decoupled from thermal bath throughout the cosmic history, it increases the effective degree of relativistic freedom, and just in the same way as extra species of massless neutrinos, its abundance is constrained by the Big Bang nucleosynthesis \cite{Starobinsky:1979ty,Tashiro:2003qp} and observation of the CMB \cite{Aghanim:2018eyx}. The effective degree of freedom induced by the graviton at the photon decoupling is quantified as \cite{Nakama2018}
\begin{equation}
	N_{GW,{\rm eff}} = \frac{4}{7} \left(\frac{4}{11}\right)^{-4/3} g_{\ast{\rm DC}} \left(\frac{g_{\ast{\rm DC}}}{g_{\ast d}}\right)^{1/3} \left.\left(\frac{\rho_{GW}}{\rho_A}\right)\right|_d \label{Neff}
\end{equation}
in terms of extra ``neutrino" generation, where $g_{\ast{\rm DC}}$ is the effective degrees of freedom at the photon decoupling and $\rho_{GW}|_d$ and $\rho_A|_d$ denote the energy density of the graviton and $A$ at the time of $A$ decay, respectively. Since $g_{\ast{\rm DC}}=3.38$ and $g_{\ast d}=106.75$, Eq.\ (\ref{Neff}) is rewritten as $N_{GW,{\rm eff}} = 2.4 (\rho_{GW}/\rho_A)|_d$. Here $\rho_{GW}|_d$ is given as \cite{Kunimitsu2012}
\begin{equation}
	\rho_{GW}|_d \simeq \frac{9H_{\rm inf}^4}{16\pi^2} a_d^{-4}. \label{rhoGW}
\end{equation}
According to Ref.\ \cite{Aghanim:2018eyx}, $N_{GW,{\rm eff}}$ must be less than 0.72, and thus Eqs.\ (\ref{rho}), (\ref{ad}), (\ref{Neff}) and (\ref{rhoGW}) yield a constraint
\begin{equation}
	\alpha^{-1/3} e^{-4m_A\Delta t} \left(\frac{m_A}{H_{\rm inf}}\right)^{5/3} > 2.3 \times 10^3. \label{constraint}
\end{equation}
In most of relevant inflationary models, the transition from inflation to kination takes place around the Hubble time. If $\Delta t = 1.0H_{\rm inf}^{-1}$, and then the left hand side of Eq.\ (\ref{constraint}) has a maximum value $4.4 \times 10^{-2}\: \alpha^{-1/3}$ at $m_A \simeq 0.42 H_{\rm inf}$. Hence, $\alpha < 7.0 \times 10^{-15}$ is required in this condition. For the cases of $k$-inflation \cite{Picon1999} and kinetically driven G-inflation \cite{Kobayashi2010}, $\Delta t$ varies from $1.2$ to $1.4H_{\rm inf}^{-1}$ and the maximum value of $\alpha$ from $1.3 \times 10^{-15}$ to $2.7 \times 10^{-15}$. When $\alpha$ takes this maximum value $\mathcal{O}(10^{-15})$, $A$ decays just before the universe turns into radiation domination Eq.\ (\ref{aMD}), and the reheating temperature reaches the order of $10^7$ GeV \cite{Hashiba2018}. The smallness of $\alpha$ can be also explained by Planckian interactions. For example, if a Yukawa-interaction is Planck-suppressed as $\tilde{\lambda} (m_A/M_{\rm Pl}) A \bar{\Psi}\Psi$, the upper bound of the Yukawa coupling $\tilde{\lambda}$ becomes just an order of unity.

\begin{figure}[tbp]
\centering
\includegraphics[width=.60\textwidth]{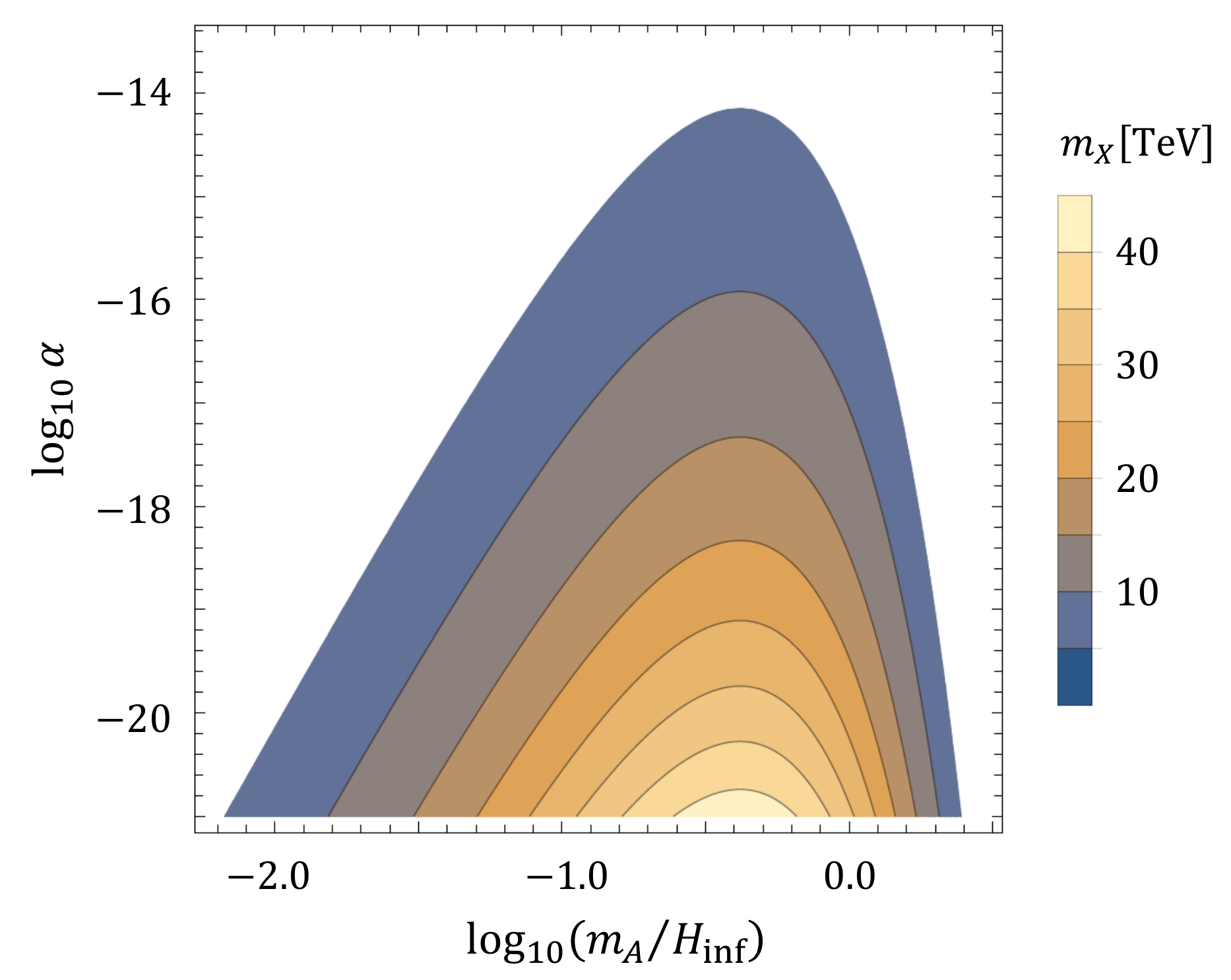}
\caption{\label{fig:mAlA} Parameter values realizing the appropriate abundance of CDM while concealing the effect of gravitationally produced gravitons with $H_{\rm inf} = 10^{13}$GeV and $\Delta t = 1.0H_{\rm inf}^{-1}$. The colored region is consistent with the CMB observation \cite{Aghanim:2018eyx}. The maximum allowed value of $\alpha$ is $7.0 \times 10^{-15}$ with $m_A = 0.42 H_{\rm inf}$. The minimum value of $m_X$ is 5.8 TeV on the edge of the allowed region.}
\end{figure}

The parameter region allowed by Eq.\ (\ref{constraint}) is shown in Fig. \ref{fig:mAlA} assuming $\Delta t = 1.0H_{\rm inf}^{-1}$. There we have also depicted contours of $m_X$ which realize proper abundance of CDM based on Eq.\ (\ref{rhos}). $m_X$ takes the minimum value $5.8\,$TeV on the edge of the allowed region. Although a lower $\alpha$ can make $m_X$ larger, it makes $T_d$ smaller much more quickly at the same time since $m_X \propto \alpha^{-1/8}$ (Eq.\ (\ref{rhos})) and $T_d \propto \alpha^{1/4}$ (Eq.\ (\ref{Td})). Therefore, $\alpha$ should be around the maximum value in order to sustain a sufficiently high reheating temperature. We assume that $m_X = \mathcal{O}(10^3)\,$GeV hereafter. In terms of the Planck-suppressed Yukawa coupling $\tilde{\lambda}$, this assumption means $\tilde{\lambda} \sim \mathcal{O}(1)$.

\section{\label{fi}Comparison with freeze-in process}
It has been claimed that PGDM may also be created by the ``freeze-in" process \cite{Hall2010,Garny2016}. In this section, we compare the amount of gravitationally produced particles and that of particles produced by the freeze-in process.

Since the interaction between PGDM and other particles is so weak that PGDM never reaches thermal equilibrium, it is created at the very high energy scale just after inflation and no longer created nor annihilated after a while. This is the freeze-in process. The effective Boltzmann equation is given in Ref.\ \cite{Giudice2001} as
\begin{align}
	\frac{dn_X}{dt} &= -3Hn_X - \langle \sigma v \rangle [n_X^2 - (n_X^{\rm eq})^2] \nonumber \\
	&\approx \langle \sigma v \rangle (n_X^{\rm eq})^2 \label{nxdt}
\end{align}
where $n_X$ is the number density of $X$, $H$ is the Hubble parameter, $n_X^{\rm eq}$ is the thermal equilibrium number density of $X$ and $\langle \sigma v \rangle$ is the thermally averaged annihilation cross section. The last line comes from $n_X \ll n_X^{\rm eq}$. $\langle \sigma v \rangle$ is given in Ref.\ \cite{Garny2016} as
\begin{equation}
	\langle \sigma v \rangle = \frac{\pi m_X^2}{M_{\rm Pl}^4} \left[ \frac{3}{5}\frac{K_1^2(z)}{K_2^2(z)} + \frac{2}{5} + \frac{4}{5}\frac{K_1(z)}{K_2(z)}z^{-1} + \frac{8}{5}z^{-2} \right],
\end{equation}
where $z \equiv m_X/T$ is a non-dimensional parameter. In our situation, $m_X \ll T_d$ and then
\begin{equation}
	\langle \sigma v \rangle \approx \frac{8\pi T^2}{5M_{\rm Pl}^4}. \label{cross}
\end{equation}
The temperature at time $t$ is derived from the energy density of radiation. Here, the energy density of $A$ and radiation density, $\rho_r = \frac{\pi^2}{30}g_\ast T^4$, obey the following Boltzmann equations:
\begin{align}
	\frac{d\rho_A}{dt} &= -3H(t)\rho_A(t) - \Gamma \rho_A(t), \label{BoltzA} \\
	\frac{d\rho_r}{dt} &= -4H(t)\rho_r(t) + \Gamma \rho_A(t). \label{Boltzr}
\end{align}
The solution of Eqs.\ (\ref{BoltzA}) and (\ref{Boltzr}) is
\begin{equation}
	\rho_r(t) = a^{-4}(t) \int_{t_f}^t dt'\, \Gamma e^{-\Gamma(t'-t_f)} a(t') \rho_A(t_f), \label{rhor}
\end{equation}
where $\rho_A(t_f)$ is given by Eq.\ (\ref{rho}). Since $m_X \ll T_d$ and the freeze-in process is effective only when the temperature is very high, the thermal equilibrium number density of $X$ is well approximated by $n_X^{\rm eq} \approx T^3/\pi^2$. Therefore, Eqs.\ (\ref{nxdt}), (\ref{cross}) and (\ref{rhor}) give the number density of $X$ produced by freeze-in process in one Hubble time around $t\ (< t_d)$ as
\begin{align}
	\frac{dn_X}{dt}H^{-1}(t) &\approx 9.4\times10^{-13} e^{-8m_A \Delta t} \frac{\Gamma^2 m_A^4 H_{\rm inf}^3}{M_{\rm Pl}^4} \nonumber \\
	&\quad\ \times \frac{\left( t^{4/3} - t_f^{4/3} \right)^2}{t_f^{2/3}} \left(\frac{t}{t_f}\right)^{-5/3},
\end{align}
which is almost proportional to $t$. This means that the freeze-in process becomes more and more efficient until $t = t_d$. Moreover, particles produced earlier are diluted by $a^{-3}$, and hence, it is enough to consider the freeze-in process around $t = t_d$. According to Eqs.\ (\ref{ad}), (\ref{nxdt}) and (\ref{cross}) the number density of particles produced by the freeze-in process in one Hubble time around $t = t_d$ is
\begin{align}
	n_X|_d &\approx \left.\left( \frac{dn_X}{dt} H^{-1} \right)\right|_d \nonumber \\
	&= \frac{8\pi T_d^2}{5\alpha M_{\rm Pl}^4 m_A} (n_X^{\rm eq})^2_d \nonumber \\
	&= 2.0\times10^{-12} \frac{\alpha m_A^5 H_{\rm inf}^2}{M_{\rm Pl}^4 e^{8m_A \Delta t}}.
\end{align}
The last line comes from Eq.\ (\ref{Td}). The produced $X$'s have the energy around $T_d$. If $\alpha$ takes its maximum value $7.0 \times 10^{-15}$, then the number density of $X$ produced by the freeze-in process is $n_X = 2.0 \times 10^{-8}\,{\rm GeV}^3$ at $t = t_d$. On the other hand, the number density of gravitationally produced particles is derived from Eq.\ (\ref{rhodecay}) as
\begin{equation}
	n_X|_d \simeq \frac{\rho_X}{m_X} = 8.1\times10^{11}\,{\rm GeV}^3.
\end{equation}
Therefore, the freeze-in process is completely negligible compared with the gravitational particle creation.

\section{\label{concl}Summary and discussion}
We have considered two scalar fields and their gravitational creation in the inflationary model where kination follows inflation such as Refs.\ \cite{Picon1999,Kobayashi2010,Peebles1999}. One of these scalar species ($A$) decays into radiation and the other ($X$) is stable and remains as the PGDM. As a result, we have found the gravitational particle creation can explain sufficient reheating and the present dark matter abundance at the same time with $H_{\rm inf} \sim 10^{13}\,$GeV, $m_A \sim 10^{13}\,$GeV, $\alpha \sim 10^{-15}$ and $m_X \sim 10^3\,$GeV.

This mass range is strongly constrained for WIMPs \cite{XENON2018}. PGDM, however, interacts with SM particles so weakly that it can escape from even the most stringent constraint. On the other hand, its feeble interaction makes it difficult to detect it experimentally. We could not help but rely on cosmological observations. One possibility is the size of dark matter clumps. Since PGDM has never reached kinetic equilibrium, it forms extremely small-scale clumps. The minimum size of the clump is typically determined by the comoving free streaming scale at the matter-radiation equality as \cite{Schneider2013}
\begin{equation}
	L_{\rm fs,eq} = \int_{t_d}^{t_{\rm eq}} dt \frac{v(t)}{a(t)} \simeq (H_{\rm inf}a^{-3}(t_{\rm RH}))^{-1} \ln\left(\frac{T_{\rm RH}}{T_{\rm eq}}\right)^2,
\end{equation}
where subscript ``eq" denotes the value at the matter-radiation equality. Therefore, the minimum mass of clump is $M_{\rm min} \sim L_{\rm fs}^3 \Omega_m \rho_{cr} \sim 2 \times 10^{-16}\,$eV. This is even much smaller than the mass of PGDM itself, and then it means that PGDM can form any size of clumps down to a few particles. Pulsar timing array observation can detect very small-scale clumps of dark matter with masses $\sim 10^{-11}$ -- $10^{-8} M_\odot$ \cite{Kashiyama2018} and then if it detects a continuous spectrum down to too small scale even for WIMPs or PBH to form, it reveals the existence of very feebly interacting dark matter --- PGDM. Of course, we can also derive the constraints on Planckian interactions of dark matters from several observations such as gamma-ray, cosmic-ray, neutrino and CMB \cite{Mambrini:2015sia}.

Finally, we comment on the effect of the non-conformal coupling. The non-conformal coupling also enhances the gravitational particle creation. In case of the minimally coupled massive scalar field, we have found that the energy density produced by the gravitational particle creation increases only by two orders of magnitude times when $m = 0.42H_{\rm inf}$ compared with the conformally coupled case. Hence, it does not change the situation dramatically.

\textit{Acknowledgements.}
We acknowledge useful comments of Toyokazu Sekiguchi and an anonymous referee. SH was supported by the Advanced Leading Graduate Course for Photon Science (ALPS). The work of JY was supported by JSPS KAKENHI, Grant JP15H02082 and Grant on Innovative Areas JP15H05888.

\bibliographystyle{apsrev4-1}
\bibliography{kination_DM}

\end{document}